\documentclass[%
 reprint,
superscriptaddress,
amsmath,amssymb,
aps,
]{revtex4-1}

\usepackage{fullpage}

\usepackage{subcaption}
\usepackage{multirow}
\usepackage{graphicx}% Include figure files
\usepackage{dcolumn}% Align table columns on decimal point
\usepackage{bm}% bold math

\begin{document}

%\preprint{APS/123-QED}

\title{Axion dark matter search using the storage ring EDM method}% Force line breaks with \\
%%\thanks{A footnote to the article title}%

\author{Seung Pyo Chang}
\affiliation{Department of Physics,KAIST,Daejeon 34141, Republic of Korea}
\affiliation{Center for Axion and Precision Physics Research, IBS, Daejeon 34051, Republic of Korea}
\author{Selcuk Haciomeroglu}%
\affiliation{Center for Axion and Precision Physics Research, IBS, Daejeon 34051, Republic of Korea}
\author{On Kim}%
\affiliation{Department of Physics,KAIST,Daejeon 34141, Republic of Korea}
\affiliation{Center for Axion and Precision Physics Research, IBS, Daejeon 34051, Republic of Korea}
\author{Soohyung Lee}%
\affiliation{Center for Axion and Precision Physics Research, IBS, Daejeon 34051, Republic of Korea}
\author{Seongtae Park}
\thanks{corresponding author}
\email{stpark@ibs.re.kr}
\affiliation{Center for Axion and Precision Physics Research, IBS, Daejeon 34051, Republic of Korea}
\author{Yannis K. Semertzidis}%
\affiliation{Department of Physics,KAIST,Daejeon 34141, Republic of Korea}
\affiliation{Center for Axion and Precision Physics Research, IBS, Daejeon 34051, Republic of Korea}


%\affiliation{
%\affiliation{
%\affiliation{%


\date{\today}% It is always \today, today,

\begin{abstract}
We propose using the storage ring EDM method to search for the axion dark matter induced EDM oscillation in nucleons. The method uses a combination of B and E-fields to produce a resonance between the $g-2$ spin precession frequency and the background axion field oscillation to greatly enhance sensitivity to it.  An axion frequency range from $10^{-9}$ Hz to 100 MHz can in principle be scanned with high sensitivity, corresponding to an $f_a$ range of $10^{13} $ GeV $\leq f_a \leq 10^{30}$ GeV,  the breakdown scale of the global symmetry generating the axion or axion like particles (ALPs).
%\begin{description}
%\end{description}
\end{abstract}

\pacs{Valid PACS appear here}% PACS, the Physics and Astronomy
\maketitle

%\tableofcontents

\section{\label{sec:level1}Introduction}

Peccei and Quinn proposed a dynamic oscillating field to solve the strong CP problem \cite{PhysRevLett.38.1440} and that oscillating field is called an axion \cite{PhysRevLett.40.223,PhysRevLett.40.279,PhysRevLett.43.103,SHIFMAN1980493,Zhitnitsky:1980tq,DINE1981199,JihnEKim}. An axion in the parameter range of $10^{11}$ GeV $\leq f_a \leq 10^{13}$ GeV is potentially observable using microwave cavity resonators, where $f_a$ is the global symmetry breakdown scale \cite{1doi:10.1146/annurev-nucl-102014-022120,2PhysRevLett.51.1415,3PhysRevD.84.055013,4PhysRevD.88.035023}. This method detects photons from the axion dark matter conversion in the presence of strong magnetic fields  \cite{2PhysRevLett.51.1415,PhysRevD.96.061102,JEONG2018412,JEONG201833,PhysRevLett.120.151301}. In the next decade it is expected that the axion frequency range of 0.1-50 GHz may be covered using microwave cavity and/or open cavity resonators \cite{yannisTalk}. However, this method cannot be used for higher values of $f_a$ (lower mass region) because the axion-photon coupling is suppressed by $f_a$ $\left(\sim 1/f_a^2\right)$ and the required resonance structures would be impractically large. For the higher values of the scale, including $M_{\rm{GUT}}$ ($\sim 10^{16}$ GeV) - $M_{\rm{PL}}$ ($\sim 10^{19}$ GeV), axion-gluon coupling can be considered, which gives a time varying electric dipole moment (EDM) to nucleons \cite{3PhysRevD.84.055013,4PhysRevD.88.035023}. For example, in the nucleon case, the EDM can be expressed as
\begin{equation}\label{eq1}
d_n = 2.4 \times 10^{-16} \frac{a}{f_a} \sim (9\times10^{-35}) \cos(m_a t) \quad \left[ e \cdot\rm{cm}\right],
\end{equation}
\begin{equation}\label{eq2}
a(t) = a_0 \cos(m_a t),
\end{equation}
where $a(t)$ is the axion dark matter field and $m_a$ is the axion mass. Graham and Rajendran proposed a method that measures the small energy shift with the form $\vec{E}\cdot\vec{d}_n$ in an atom as a probe of the oscillating axion field \cite{3PhysRevD.84.055013}. In this case, the electric field is an internal atomic field.
By combining Eq. (\ref{eq1}) and Eq. (\ref{eq2}) with a possible static EDM, one can write the total EDM as,
\begin{equation}\label{eq3}
d(t)=d_{DC}+d_{AC}\cos(m_a t+\varphi_{ax}),
\end{equation}
where $d_{DC}$ and $d_{AC}$ are the magnitudes of the static and oscillating parts of EDM, respectively, and $\varphi_{ax}$ is the phase of the axion field. In this paper, we propose using the storage ring technique to probe the oscillating EDM signal \cite{protonEDMproposalToBNL,deuteronEDMproposalToBNL,5doi:10.1063/1.4967465}, with some modification of storage ring conditions depending on the axion frequency. Instead of completely zeroing the $g-2$ frequency, we just control and tune it to be in resonance with the axion background field oscillation frequency. We propose searching for the oscillating EDM term by using a resonance with the $g-2$ precession frequency. This method is expected to be more sensitive, and the systematic errors are easier to handle than in the frozen spin storage ring EDM method. Using the storage ring method, one can scan the frequency range from $10^{-9}$ Hz up to 100 MHz, which corresponds to an axion parameter space of about $10^{13} $ GeV $\leq f_a \leq 10^{30}$ GeV.

\section{\label{sec:level2}Resonance of Axion induced oscillating EDM with $g-2$ spin precession in storage rings}
The previously proposed storage ring EDM experiment is optimized for a DC (fixed in time) nucleon EDM, applied to protons and deuterons \cite{protonEDMproposalToBNL,deuteronEDMproposalToBNL,5doi:10.1063/1.4967465}. It is designed to keep (freeze) the particle spin direction along the momentum direction for the duration of the storage time, typically for $10^3$ s, the stored beam polarization coherence time \cite{5doi:10.1063/1.4967465,PhysRevLett.117.054801}. In this case, the radial electric field in the particle rest frame is precessing the particle spin in the vertical plane. The precession frequency in the presence of both E and B fields is expressed by the T-BMT equation Eq. (\ref{eq4})-(\ref{eq6}) \cite{6PhysRevLett.2.435,7jackson_classical_1999}.

\begin{equation}\label{eq4}
\vec{\omega}=\vec{\omega}_a+\vec{\omega}_d,
\end{equation}

\begin{equation}\label{eq5}
\vec{\omega}_a=-\frac{e}{m}\left[a\vec{B}-\left(a-\frac{1}{\gamma^2-1}\right)\frac{\vec{\beta}\times\vec{E}}{c}\right],
\end{equation}

\begin{equation}\label{eq6}
\vec{\omega}_d=-\frac{e}{m}\left[\frac{\eta}{2}\left(\frac{\vec{E}}{c}+\vec{\beta}\times\vec{B}\right)\right],
\end{equation}
where $a=\left(g-2\right)/2$ is the magnetic anomaly with $a=-0.14$ for deuterons. Here, other terms are omitted by assuming the conditions $\vec{\beta} \cdot \vec{E}=\vec{\beta} \cdot \vec{B}=0$. The parameter $\eta$ shown in the equation is related to the electric dipole moment $d$ as $d=\eta e \hbar / 4mc$. Since we are dealing with a time varying EDM due to the oscillating axion background field, $\eta$ is also a function of time. The $\vec{\omega}_a$ is the angular frequency, i.e., 2$\pi$ times the $g-2$ frequency, describing the spin precession in the horizontal plane relative to the momentum precession.

The term $\vec{\omega}_d$ is due to the EDM and the corresponding precession takes place in the vertical plane. For a time independent nucleon EDM, the spin vector will precess vertically for the duration of the storage time if the horizontal spin component is fixed to the momentum direction \cite{5doi:10.1063/1.4967465}. This condition can be achieved by setting the E and B fields properly, and is called the frozen spin condition.

With a nonzero $g-2$ frequency, the average EDM precession angle becomes zero for the static EDM case because the relative E field direction to the spin vector changes within every cycle of $g-2$ precession. For example, the spin tilts in one direction (up or down) due to the EDM within one half cycle and then tilts in the opposite direction for the other half cycle, resulting in an average accumulation of zero. The presence of a static EDM will only slightly tilt the $g-2$ precession plane away from the horizontal plane, without a vertical spin accumulation. In contrast, for an oscillating EDM, when the axion frequency ($\omega_{ax}$) is the same as the $g-2$ frequency with the appropriate phase, the precession angle can be accumulated in one direction. This is possible because the EDM direction flips every half cycle due to the axion filed oscillation and the relative direction between the E field and the EDM $d$ always remains the same. 

In this idea of resonant axion induced EDM with $g-2$ spin precession, one can utilize the strong effective electric field $\vec{E^*}=\vec{E}+c \vec{\beta} \times \vec{B}$, which comes from the B field due to particle motion, as expressed in Eq. (\ref{eq6}). In this case, the effective electric field is about one or two orders of magnitude larger than the applied external E field which can be up to 10 MV/m and has an apparent technical limitation in strength.

\section{Sensitivity calculation}
The statistical error in the EDM for proton or deuteron can be expressed with the following equation \cite{5doi:10.1063/1.4967465,deuteronEDMproposalToBNL}.
\begin{equation}\label{eq7}
\sigma_d = \frac{2\hbar s}{PAE^*\sqrt{N_c\kappa T_{tot} \tau_p}},
\end{equation}
where $P$ is the degree of polarization, $A$ the analyzing power, $E^*$ the effective electric field that causes the EDM precession, $N_c$ the number of particles stored per cycle, $\kappa$ the detection efficiency of the polarimeter, $\tau_p$ is the polarization life time, and $T_{tot}$ the total experiment running time. The s in the numerator is 1/2 for protons and 1 for deuterons.  One can easily calculate the sensitivity of nucleon EDM measurement using this formula with the corresponding experimental parameters.

In this study, we used the following method to calculate the sensitivities of the axion EDM measurement including the oscillation effect. First, we chose the target axion frequency and then calculated the corresponding E and B-fields for the particle storage, which give the same $g-2$ frequency as the chosen axion frequency. Then, the axion oscillation and $g-2$ spin precession will be on resonance and the EDM precession angle in the vertical plane can keep accumulating during the measurement time. With the chosen axion frequency and the axion quality factor $Q_{ax}$, we estimated the statistical error and the resulting error was used to calculate the experiment sensitivity along with the effective electric field. This method can be used not only for nucleons like deuterons or protons but it can also be used for other leptonic particles like muons, provided there is a coupling between the oscillating $\theta_{QCD}$ induced by the background axion dark matter field and the particle EDM.

As shown in Eq. (\ref{eq6}), the EDM part of the precession rate can be rewritten as Eq. (\ref{eq8})
\begin{equation}\label{eq8}
\left.\begin{aligned}
\omega_d=\frac{d\theta}{dt}=-\frac{d}{s \hbar}E^*,\\ E^* = E+c\beta B,
\end{aligned}\right.
\end{equation}
where s is 1/2 for protons and 1 for deuterons. Accordingly, $d=\frac{s\hbar}{E^*}\omega_d$, and the error for the EDM $d$ can be written as Eq.  (\ref{eq9}).
\begin{equation}\label{eq9}
\sigma_d=\frac{s\hbar}{E^*}\sigma_{\omega_d},
\end{equation}
where $\sigma_{\omega_d}$ is the error for $\omega_{d}$. It can be obtained from the fit of the vertical precession angle $\theta$ as a function of time. 

The time variation of $\theta(t)$ will be obtained from the asymmetry $\epsilon (t)$ measurement (explained below) using a polarimeter \cite{0034-4885-35-2-305,OHLSEN197341,BONIN1990389, PhysRevC.23.616,BRANTJES201249}. There is currently no nondestructive way to measure the spin direction of particles with large sensitivity. For the hadronic particle case, one can utilize the nuclear interactions between the spin polarized particles and target nuclei. In this case, the spin-orbit interaction between the spin polarized incident particle and target nucleus is one of the major reactions that gives asymmetrical scattering of the incident particle in the azimuthal angle \cite{MARMIER19701019}. Carbon is one of the most efficient target materials with large analyzing power for both deuterons and protons. 

For example, the scattering cross section for a spin 1/2 polarized particle (e.g., a proton) can be written as Eq. (\ref{eq10}) \citep{OHLSEN197341}
\begin{equation}\label{eq10}
I(\psi,\phi)= I_0(\psi)\left[1+p_yA_y(\psi)\right],
\end{equation}
where $I_0(\psi)$ is the cross section for an unpolarized particle scattered into the angle $\psi$, $A_y(\psi)$ is the analyzing power of the reaction, and $p_y$ is the transverse component of the beam polarization. From Eq. (\ref{eq10}), the number of hits recorded in a detector located at ($\psi$, $\phi$) can be written as follows
\begin{equation}\label{eq11}
N(\psi,\phi)=nN_A\Delta\Omega\zeta I(\psi,\phi),
\end{equation}
where $n$ is the number of particles incident on the target, $N_A$ is the number of target nuclei per square centimeter, $\Delta\Omega$ is the solid angle subtended by the detector and $\zeta$ is the efficiency of the detector. If we assume identical detectors which are placed at symmetrical locations on the left and right of the beam direction, the counts recorded on the left and right are

\begin{equation}\label{eq12}
L\equiv N(\psi,0)=n N_A \Delta\Omega\zeta I_0(\psi)[1+P_yA_y(\psi)],
\end{equation}
\begin{equation}\label{eq13}
R\equiv N(\psi,0)=n N_A \Delta\Omega\zeta I_0(\psi)[1-P_yA_y(\psi)],
\end{equation}
respectively. After simple algebra one can obtain the left-right asymmetry $\epsilon$ for the vertically polarized beam as follows
\begin{equation}\label{eq14}
\epsilon (t)=\frac{L-R}{L+R}(t)=PA\theta(t),
\end{equation}
where $A$ is the analyzing power, $\theta(t)$ is the accumulated EDM precession angle in the vertical plane and $L$ and $R$ are the number of hits on the left and right detectors, respectively. As mentioned earlier, one can get the EDM precession rate $\omega_d$ from the asymmetry $\epsilon (t)$.

For error estimation, we assumed an axion quality factor of $Q_{ax}=3\times 10^6$. From the $Q_{ax}$ value, one can calculate the possible measurement time $t_m=Q_{ax}/f_{ax}$ (coherence time between the axion oscillation and $g-2$ spin precession), where $f_{ax}$ is the axion frequency \cite{3PhysRevD.84.055013}. If this time $t_m$ is larger than the polarization life time $t_{pol}$ (or spin coherence time, SCT), then we set the measurement time to $t_{pol}$. Otherwise, the measurement time is set to $Q_{ax}/f_{ax}$. Furthermore, in some models the oscillating axion field is monochromatic with a quality factor in excess of $10^{10}$; see references \cite{9Sikivie:2010fa,SIKIVIE20031} and references therein. As shown in Table \ref{table1}, with this high $Q_{ax}$ value one can reach very high sensitivities up to $10^{-31\sim -32}~e\cdot$cm under the said assumptions. 

In resonance microwave cavity experiments, the conversion power of axion to photon is limited by the cavity quality factor $Q_L$. Therefore, there is no large benefit from the high axion $Q_{ax}$ value if the cavity $Q_L$ is smaller than the axion $Q_{ax}$ value. The current cavity experiment assumes the axion $Q_{ax}$ to be about a few times $10^6$ and the cavity $Q_L$ is usually smaller than this. However, the proposed storage ring method can obtain a large benefit from the large $Q_{ax}$ value since the spin tune stability in the storage ring becomes very large and the sensitivity is even more enhanced with higher $Q_{ax}$ values, as shown in Tables \ref{table1},\ref{table2}. For example, it was experimentally measured at COSY that the spin tune was controllable at the precision level of $10^{-10}$ for a continuous $10^2$ s accelerator cycle time \cite{10PhysRevLett.115.094801}.

\subsection{Pure magnetic ring}
\begin{figure}
\includegraphics[width=\linewidth]{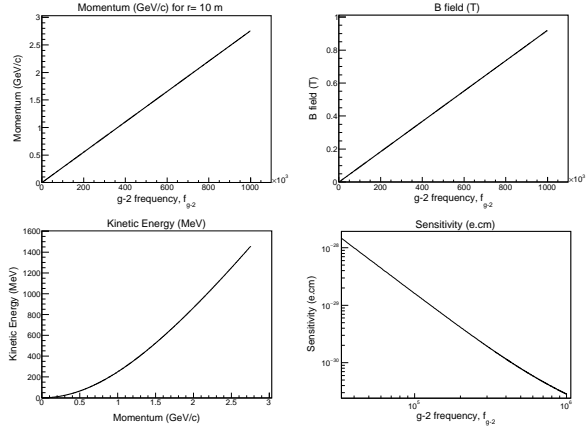}
\caption{Deuteron sensitivity vs. $g-2$ frequency (or axion frequency). Purely magnetic ring only with ring bending radius $r=10$ m.}
\label{fig1}
\end{figure}

The effective electric field is an important parameter in the sensitivity estimation. As shown in Eq. (\ref{eq6}), the $\vec{E}^*$ is the vector sum of the radial $\vec{E}$ field and $\vec{v}\times\vec{B}$. For this study, we started with a pure magnetic ring first. We assumed the ring bending radius to be $r=10~{\rm m}$. In order to tune the $g-2$ frequency $ f_{\rm g-2}$ (or axion frequency on the resonance condition), the B field was varied and the momentum was also changed accordingly to keep the ring bending radius unchanged. The bottom right plot in Fig. \ref{fig1} shows the sensitivity as a function of the $g-2$ frequency and Eq. (\ref{eq7}) is used for the calculation. As can be seen in the plot, the experiment is more sensitive at high frequencies. This is because the larger B field provides a larger effective E field by $\vec{v}\times\vec{B}$. Below $\sim 10^5$ Hz, the sensitivity decreases beyond $\geq 10^{-29}~e\cdot$ cm. We decided to use the  E and B field combined ring for the low frequency region to improve the sensitivity in that range.

\subsection{E/B combined ring}
\begin{figure}
\centering
\begin{subfigure}{.4\textwidth}
  \centering
  \includegraphics[width=0.8\linewidth]{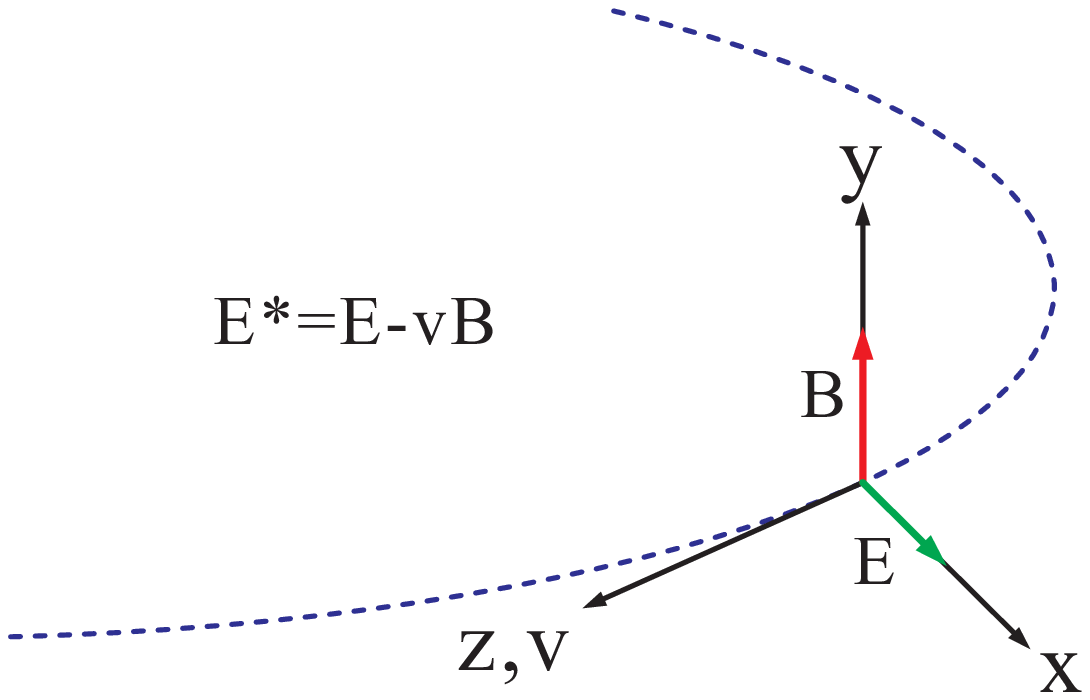}
  \caption{The coordinate of B, E and $v$}
  \label{fig2a}
\end{subfigure}%
\hfill
\begin{subfigure}{.4\textwidth}
  \centering
  \includegraphics[width=1\linewidth]{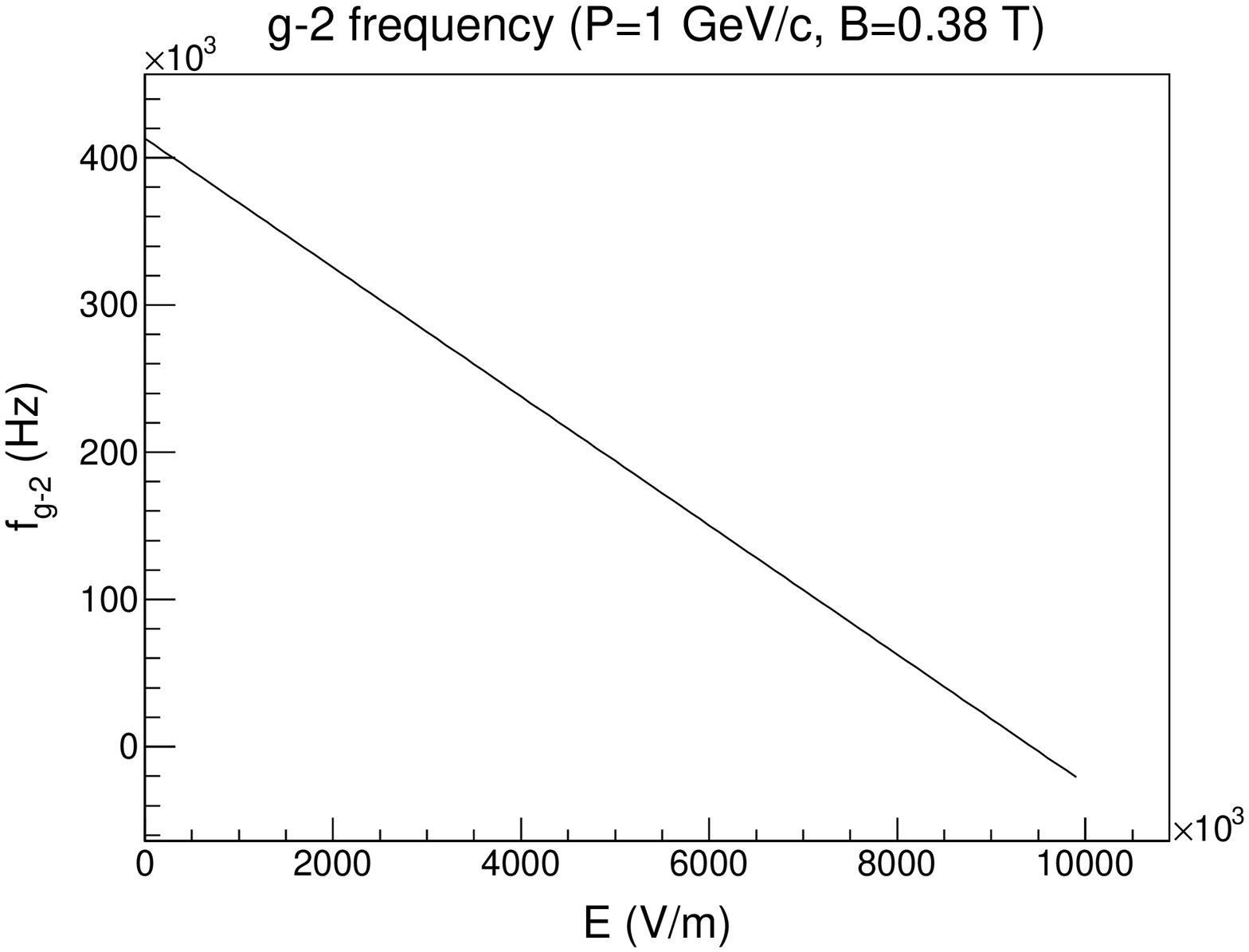}
  \caption{$g-2$ frequency vs E field}
  \label{fig2b}
\end{subfigure}
\caption{E/B combined ring for $g-2$ frequency tunning}
\label{fig2}
\end{figure}

The B field was set to 0.38 T and the E field was applied to the radial direction (radially outwards indicates positive direction) as shown in Fig. \ref{fig2a}. Fig. \ref{fig2b} shows the $g-2$ frequency as a function of the applied E field. To increase the frequency, the E field has to be reduced. However, with this E field change, the ring radius changes as well. In order to keep the ring radius unchanged ($r=10~{\rm m}$ for deuteron in this example), the momentum is adjusted accordingly. Some examples of the experimental conditions are listed in Table \ref{table1}. 

Using the parameters shown in Table \ref{table1}, we generated the asymmetry data and the data was fit to a function which is a combination of a linear function and an exponential function reflecting polarization decay. From the fit, we obtained the error for the precession frequency, $\sigma_{\omega_d}$. The fit error was inserted into Eq. (\ref{eq9}) to calculate the error of the EDM $d$. 

\begin{figure}
\includegraphics[width=\linewidth]{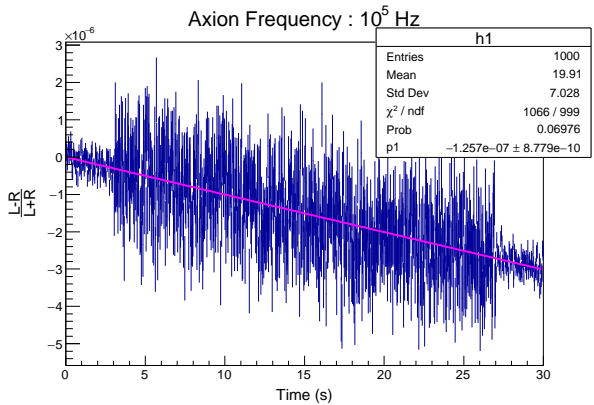}
\caption{Simulation result of asymmetry vs. time with fit.}
\label{fig3}
\end{figure}

An example of the simulation result with the fit is shown in Fig. \ref{fig3}. In the simulation, 40$\%$ of the particles were extracted in the early 10$\%$ of measurement time (or storage time) and another 40$\%$ was extracted in the late 10$\%$ of measurement time. The remaining 20$\%$ of the particles was extracted in the middle 80$\%$ measurement time. The middle 80$\%$ of time can be used to measure the g-2 frequency. The calculated sensitivities are also shown in Table \ref{table1} for the deuteron case and are all about $10^{-30}~e \cdot$cm or less.

\begin{table*}
	\caption{Examples of experimental parameters for frequency tuning, and results of sensitivity calculation (Deuteron). The analyzing power was assumed to be $A=0.36$ if the momentum P was below 2 GeV/$c$ and A=0.15 was used for the momentum P$>$2 GeV/$c$. The ring bending radius was 10 m. The polarimeter efficiency was assumed to be 2\% and initial polarization was 0.8. The axion quality factors: $Q_{ax1}=3\times10^6$, $Q_{ax2}=10^{10}$.}
	\label{table1}
	\begin{tabular}{|c|c|c|c|c|c|c|c|c|c|c|}
		\hline
		\multirow{3}{*}{B (T)} & \multirow{3}{*}{P (GeV/$c$)} & \multirow{3}{*}{$f_{g-2}$} & \multirow{3}{*}{$\rm{E}_{r}$ (V/m)} & \multirow{3}{*}{E* (V/m)} & \multicolumn{6}{c|}{Sensitivity (e$\cdot$cm)}             \\ \cline{6-11}
		&                   &                   &                   &                   & \multicolumn{2}{c|}{$SCT=10^3 s$} & \multicolumn{2}{c|}{$SCT=10^4 s$} & \multicolumn{2}{c|}{$SCT=10^5 s$} \\ \cline{6-11}
		&           &            &            &           & $Q_{ax1}$   & $Q_{ax2}$ & $Q_{ax1}$   & $Q_{ax2}$ & $Q_{ax1}$ & $Q_{ax2}$ \\ \hline
		0.3800 & 0.9428  & 1.E1 & 8.82E6  & 4.23E7 & 9.9E-31 & 9.9E-31 & 3.1E-31 & 3.1E-31 & 9.9E-32 & 9.9E-32 \\ \hline
		0.3800 & 0.9429  & 1.E2 & 8.82E6  & 4.23E7 & 9.9E-31 & 9.9E-31 & 3.1E-31 & 3.1E-31 & 1.4E-31 & 9.9E-32 \\ \hline
		0.3800 & 0.9433  & 1.E3 & 8.80E6  & 4.24E7 & 9.9E-31 & 9.9E-31 & 4.3E-31 & 3.1E-31 & 3.8E-31 & 9.9E-32 \\ \hline
		0.3800 & 0.9473  & 1.E4 & 8.65E6  & 4.27E7 & 1.4E-30 & 9.9E-31 & 8.3E-31 & 3.1E-31 & 1.2E-30 & 9.9E-32 \\ \hline
		0.3800 & 0.9880  & 1.E5 & 7.05E6  & 4.60E7 & 3.5E-30 & 9.1E-31 & 3.5E-30 & 2.9E-31 & 3.5E-30 & 9.1E-32 \\ \hline
		0.3800 & 1.0345  & 2.E5 & 5.06E6  & 5.00E7 & 4.6E-30 & 8.4E-31 & 4.5E-30 & 2.7E-31 & 4.5E-30 & 9.8E-32 \\ \hline
		0.3800 & 1.1326  & 4.E5 & 3.47E5  & 5.85E7 & 5.5E-30 & 7.2E-31 & 5.5E-30 & 2.3E-31 & 5.5E-30 & 5.2E-32 \\ \hline
		0.3800 & 1.2386  & 6.E5 & -5.47E6 & 6.82E7 & 5.8E-30 & 6.2E-31 & 5.3E-30 & 2.0E-31 & 5.8E-30 & 1.1E-31 \\ \hline
		0.3800 & 1.3546  & 8.E5 & -1.26E7 & 7.93E7 & 5.7E-30 & 5.3E-31 & 3.9E-30 & 1.7E-31 & 5.7E-30 & 1.0E-31 \\ \hline
		0.3800 & 1.4836  & 1.E6 & -2.14E7 & 9.20E7 & 5.5E-30 & 4.6E-31 & 3.5E-30 & 1.4E-31 & 5.5E-30 & 1.0E-31 \\ \hline
		0.8000 & 2.5124  & 1.E6 & -9.13E6 & 2.01E8 & 1.6E-30 & 1.5E-31 & 2.5E-30 & 6.6E-32 & 2.5E-30 & 4.6E-32 \\ \hline
		0.9198 & 2.7574  & 1.E6 & 0.0     & 2.28E8 & 5.3E-30 & 4.4E-31 & 3.4E-30 & 1.4E-31 & 5.3E-30 & 9.7E-32 \\ \hline
		9.1977 & 27.5740 & 1.E7 & 0.0     & 2.75E9 & 3.3E-30 & 3.7E-32 & 4.4E-30 & 2.5E-32 & 4.4E-30 & 2.4E-32 \\ \hline
\end{tabular}
\end{table*}

\begin{table*}
	\centering
	\caption{Examples of experimental parameters for frequency tuning, and results of sensitivity calculation (Proton). The analyzing power was assumed to be $A=0.6$ for the momentum P$<$1 GeV/$c$ and $A=0.25$ was used for the momentum P$>$1 GeV/$c$. The ring bending radius was 52 m for the E/B combined ring, and $r=10$  m for the pure magnetic ring. The polarimeter efficiency used was 2\% and initial polarization was 0.8. The axion quality factors: $Q_{ax1}=3\times10^6$, $Q_{ax2}=10^{10}$.}
	\label{table2}
	\begin{tabular}{|c|c|c|c|c|c|c|c|c|c|c|}
		\hline
		\multirow{3}{*}{B (T)} & \multirow{3}{*}{P (GeV/$c$)} & \multirow{3}{*}{$f_{g-2}$} & \multirow{3}{*}{$\rm{E}_{r}$ (V/m)} & \multirow{3}{*}{E* (V/m)} & \multicolumn{6}{c|}{Sensitivity (e$\cdot$cm)}             \\ \cline{6-11}
		&                   &                   &                   &                   & \multicolumn{2}{c|}{$SCT=10^3 s$} & \multicolumn{2}{c|}{$SCT=10^4 s$} & \multicolumn{2}{c|}{$SCT=10^5 s$} \\ \cline{6-11}
		&           &            &            &           & $Q_{ax1}$   & $Q_{ax2}$ & $Q_{ax1}$   & $Q_{ax2}$ & $Q_{ax1}$ & $Q_{ax2}$ \\ \hline
		0.00011  & 0.6984  & 1.E1 & -8.0E6 & 8.02E6 & 1.6E-30 & 1.6E-30 & 5.0E-31 & 5.0E-31 & 1.6E-31 & 1.6E-31 \\ \hline
		0.00010  & 0.6984  & 1.E2 & -8.0E6 & 8.02E6 & 1.6E-30 & 1.6E-30 & 3.6E-31 & 3.6E-31 & 2.2E-31 & 1.6E-31 \\ \hline
		0.00008  & 0.6982  & 1.E3 & -8.0E6 & 8.01E6 & 1.6E-30 & 1.6E-30 & 6.9E-31 & 5.0E-31 & 6.1E-31 & 1.6E-31 \\ \hline
		-0.00015 & 0.6960  & 1.E4 & -8.0E6 & 7.97E6 & 2.2E-30 & 1.6E-30 & 1.9E-30 & 4.1E-31 & 1.9E-30 & 1.6E-31 \\ \hline
		-0.00243 & 0.6747  & 1.E5 & -8.0E6 & 7.57E6 & 6.4E-30 & 1.7E-30 & 4.9E-30 & 5.3E-31 & 6.4E-30 & 1.7E-31 \\ \hline
		-0.00495 & 0.6519  & 2.E5 & -8.0E6 & 7.15E6 & 9.6E-30 & 1.8E-30 & 9.5E-30 & 5.6E-31 & 6.7E-30 & 2.0E-31 \\ \hline
		-0.01523 & 0.7103  & 4.E5 & -1.1E7 & 8.24E6 & 1.2E-29 & 1.1E-30 & 1.2E-29 & 4.8E-31 & 1.2E-29 & 2.3E-31 \\ \hline
		-0.02002 & 0.6711  & 6.E5 & -1.1E7 & 7.51E6 & 1.6E-29 & 1.7E-30 & 1.4E-29 & 5.3E-31 & 1.6E-29 & 2.9E-31 \\ \hline
		-0.02666 & 0.6643  & 8.E5 & -1.2E7 & 7.38E6 & 1.8E-29 & 1.7E-30 & 1.8E-29 & 5.4E-31 & 1.8E-29 & 3.4E-31 \\ \hline
		-0.03327 & 0.6583  & 1.E6 & -1.3E7 & 7.27E6 & 2.1E-29 & 1.7E-30 & 2.1E-29 & 5.5E-31 & 2.1E-29 & 3.8E-31 \\ \hline
		0.36587  & 1.0968  & 1.E7 & 0.0      & 8.33E7 & 4.4E-29 & 3.6E-31 & 4.4E-29 & 1.9E-31 & 4.4E-29 & 2.4E-31 \\ \hline
		3.65868  & 10.9684 & 1.E8 & 0.0      & 1.09E9 & 2.3E-29 & 3.9E-32 & 3.4E-29 & 5.8E-32 & 3.4E-29 & 5.8E-32 \\ \hline
	\end{tabular}
\end{table*}

The polarimeter efficiency and the average analyzing power used for the sensitivity calculation were $2 \%$ and 0.36, respectively, for the deuteron case (Table \ref{table1}). The numbers are for the elastic d-C reaction (exclusive reaction) measured at the deuteron energy of 270 MeV (p=1042  MeV/c) \cite{11SATOU2002307}. The analyzing power is a function of particle energy and one should avoid the energies that have small analyzing powers. For a frequency of 1 MHz, the required momentum is about 2.8 GeV/c ($T \sim 1.5$ GeV) for a magnetic field of 0.92 T, as shown in Table \ref{table1}. In the literature \cite{12LADYGIN1998129} one can find corresponding analyzing powers for the inclusive d-C reaction to be maximum about 0.15 within the angle range we are interested in, $5^o-20^o$. This value is still practically useful, but beyond this momentum the analyzing power might be too small to be used for the deuteron polarization analysis.

We repeated the estimations for the proton case as well and the results are shown in Table \ref{table2}. For the proton, we assumed the ring bending radius to be 52 m. This is the ring radius value suggested for the static EDM measurement by the storage ring proton EDM collaboration \cite{5doi:10.1063/1.4967465}. The proton has a larger magnetic anomaly (G=1.79) than the deuteron and its mass is half that of the deuteron. Therefore, its spin precession rate in the magnetic field is about 26 times more sensitive than the deuteron case. For this reason, small magnetic fields have to be used for a low axion frequency scan. Because of the small magnetic field (contributing to effective electric field $E^*$), the calculated sensitivities are not as good as in the deuteron case. However, the resulting sensitivities are still comparable to the static EDM case ($\sim 10^{-29}~e\cdot$cm). Instead of varying the electric field strength (as was done for the deuteron case), the magnetic field was changed to modify the $g-2$ frequency. We changed the momentum as needed to keep the same ring radius. However, we kept the proton kinetic energy at around 200 MeV ($P \geq 650$  MeV/c) to keep a large analyzing power from the polarimeter detector. The average analyzing power for a proton energy of about 200 MeV is about 0.6 \cite{13MCNAUGHTON1985435} and this value was used in the sensitivity estimation as shown in Table \ref{table2}.

The last two rows in Table \ref{table2} are for the presence of B-field only and the ring bending radius used was $r=10$ m. High frequency ($\geq 10^7$ Hz) can be easily reached at small B-fields. However, for a constant ring bending radius, the momentum has to be changed. When the proton case is combined with the deuteron results shown in Table \ref{table1}, one can perform measurements from $10^{-9}$ Hz to 100 MHz using the same storage ring with a bending radius of $r=10$ m.

\begin{figure*}
	\begin{minipage}{0.9\textwidth}
		\centering
		\begin{subfigure}[t]{0.45\textwidth}
			\centering
			\includegraphics[width=1.0\linewidth]{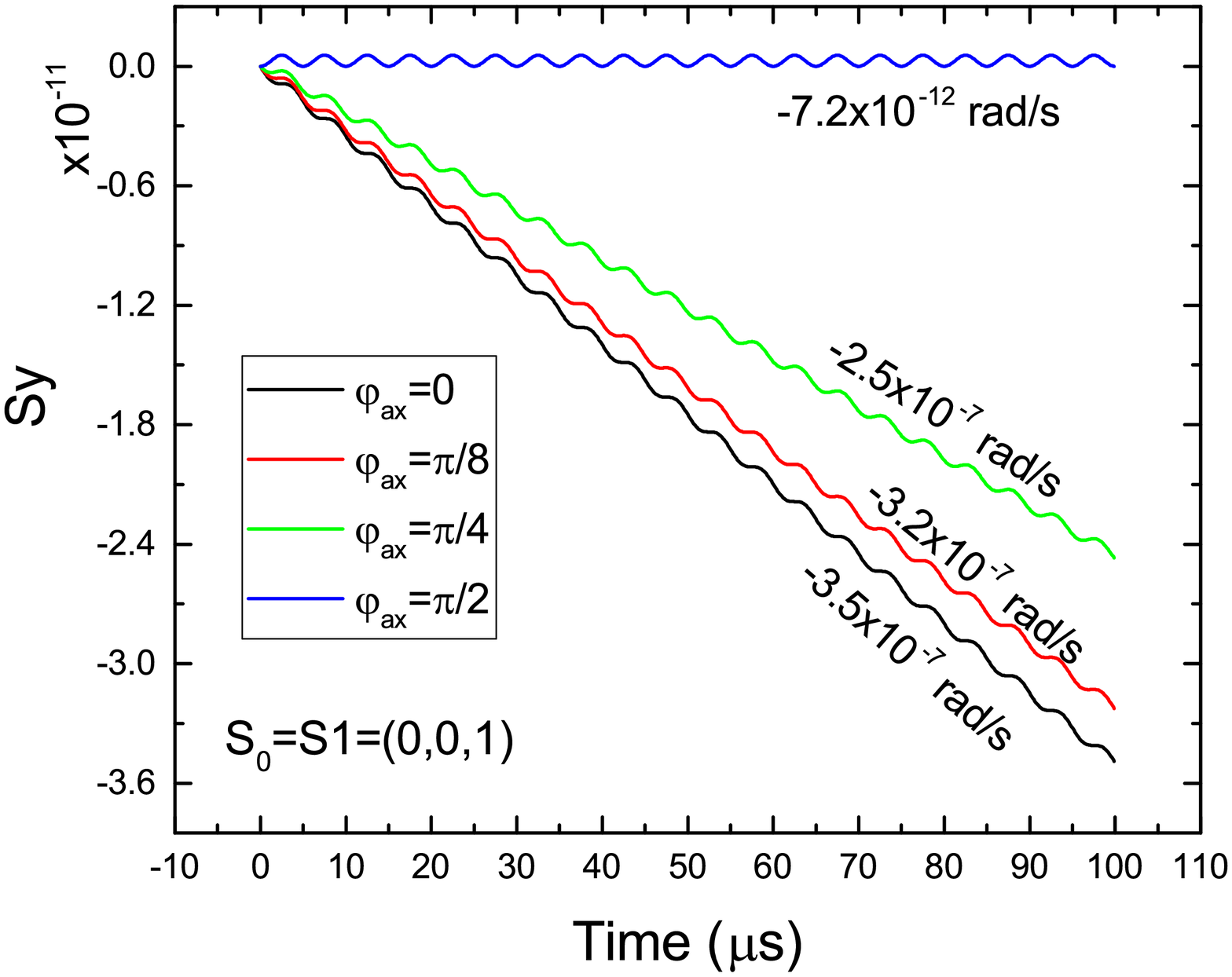} 
			\caption{Axion phase dependence of EDM precession rates. \label{fig:label:4a}}
		\end{subfigure}
%		\hfill
		\begin{subfigure}[t]{0.45\textwidth}
			\centering
			\includegraphics[width=1.0\linewidth]{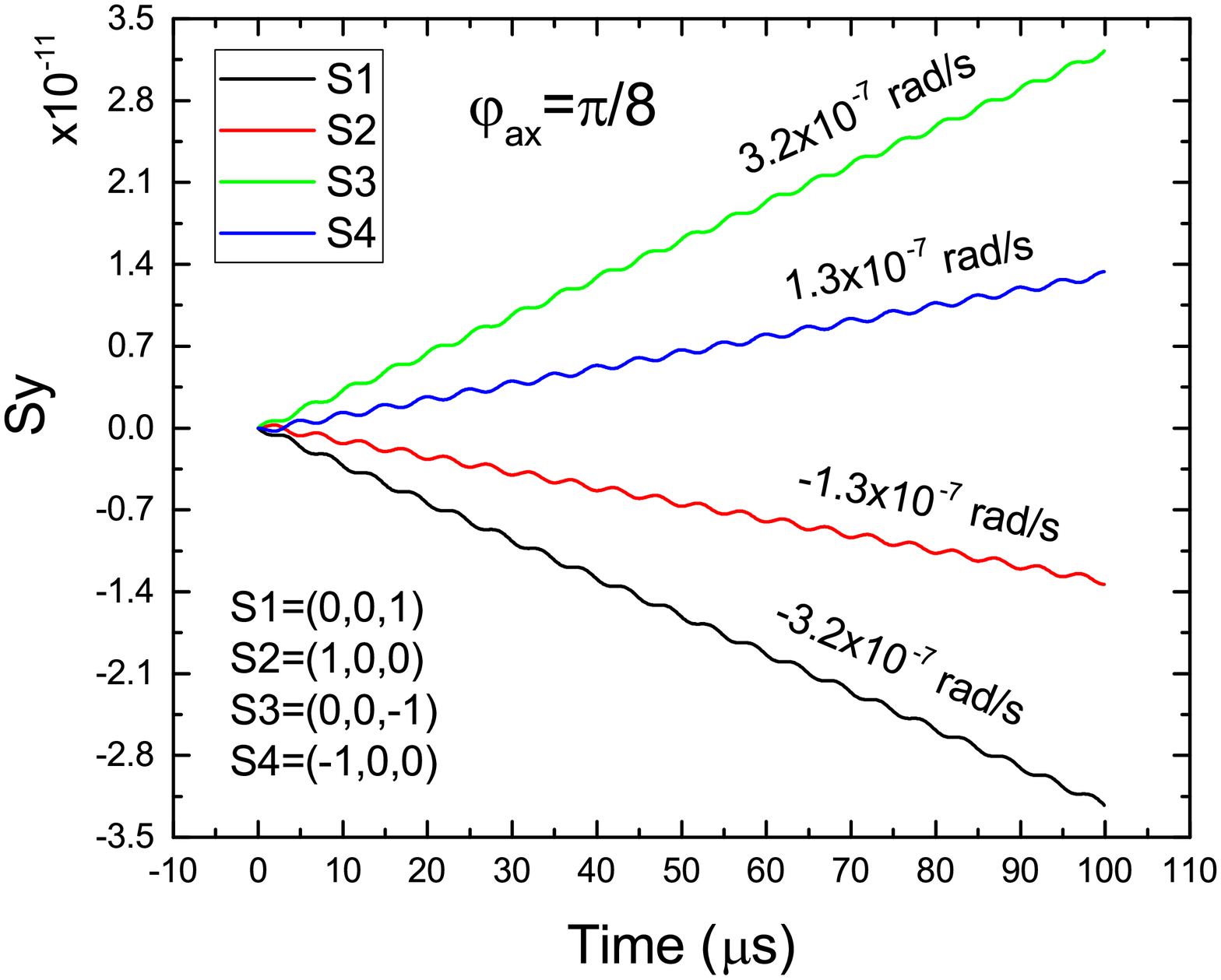} 
			\caption{Spin polarization dependence of EDM precession rates. Initial axion phase $\varphi_{ax}=\pi/8$ \label{fig:label:4b}}
		\end{subfigure}
		%    \vspace{1cm}
%		\hfill
		\begin{subfigure}[t]{0.45\textwidth}
			\centering
			\includegraphics[width=0.8\linewidth]{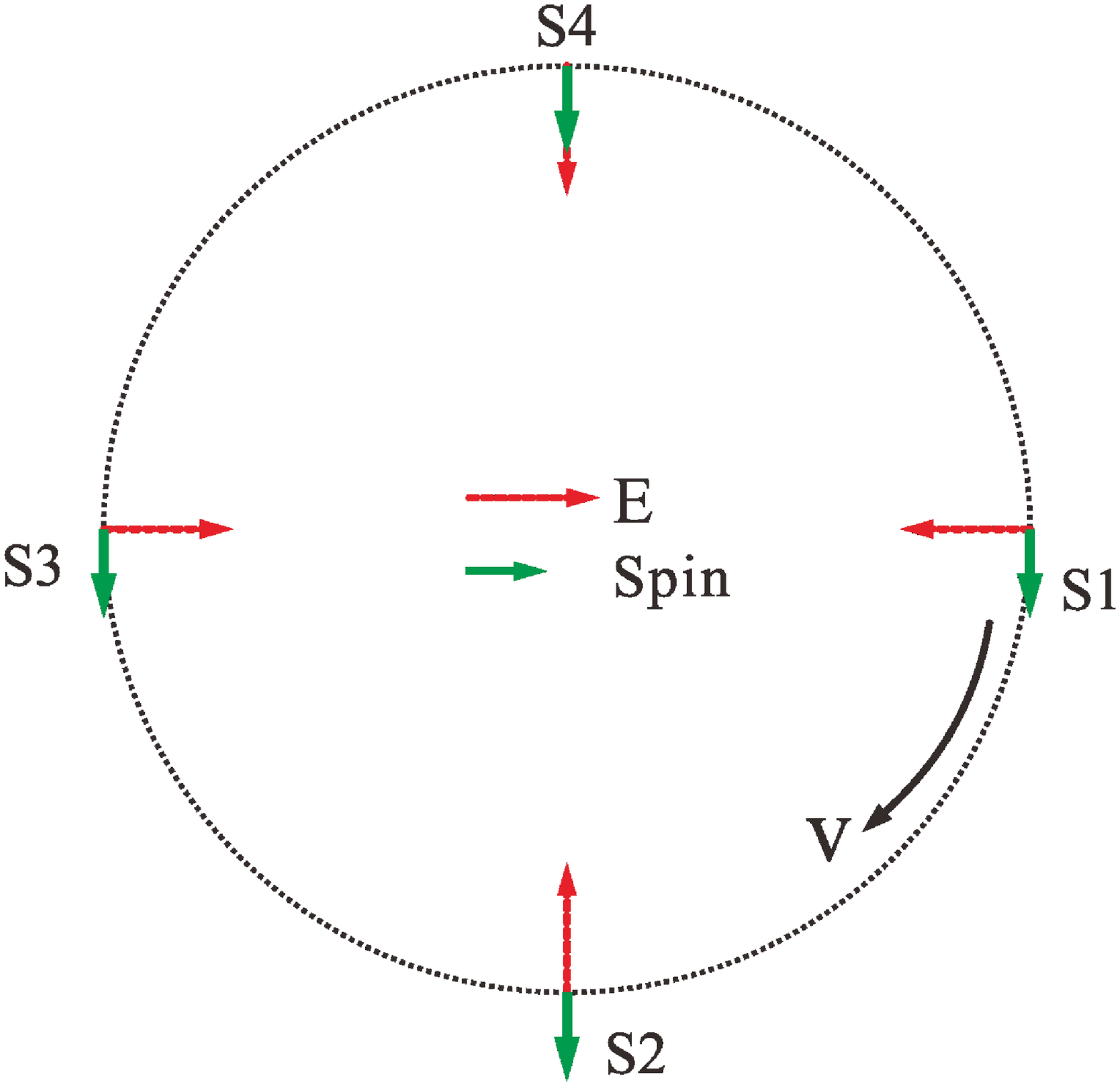} 
			\caption{Example of four orthogonal spin polarization settings. \label{fig:label:4c}}
		\end{subfigure}
		\caption{Axion phases and EDM precession rates. }\label{fig:label4}
	\end{minipage}
\end{figure*}

\section{Axion phase effect}

Since the initial phase of the axion field is unknown, the phase $\varphi_{ax}$ that appears in Eq. (\ref{eq3}) cannot be controlled in the experiment. However, the rate of the EDM precession angle strongly depends on the relative phase between the initial spin and axion phase $\varphi_{ax}$. Fig. \ref{fig:label4} shows the effect of initial phase on the vertical spin precession (EDM effect). The parameters used in the simulation are for an axion frequency of $10^5$ Hz, which is shown in Table \ref{table1}. The spin tracking was done by integrating the following two equations \cite{6PhysRevLett.2.435,7jackson_classical_1999} for spin and velocity, respectively.

\begin{equation}\label{eq15}
\begin{split}
\frac{d\vec{s}}{dt}=\frac{e}{m}\vec{s}\times\Bigg[&\left(\frac{g}{2}-\frac{\gamma-1}{\gamma}\right)\vec{B}\\&-\left(\frac{g}{2}-1\right)\frac{\gamma}{\gamma+1}\left(\vec{\beta}\cdot\vec{B}\right)\vec{\beta} \\& -\left(\frac{g}{2}-\frac{\gamma}{\gamma+1}\right)\frac{\vec{\beta}\times \vec{E}}{c} \\& +\frac{\eta}{2}\left(\vec{\beta}\times\vec{B}+\frac{\vec{E}}{c}-\frac{\gamma}{\gamma+1}\frac{\vec{\beta}\cdot\vec{E}}{c}\vec{\beta}\right)\Bigg],\\
\end{split}
\end{equation}

\begin{equation}\label{eq16}
\frac{d\vec{\beta}}{dt}=\frac{e}{\gamma m}\left[\vec{\beta}\times\vec{B}+\frac{\vec{E}}{c}-\frac{\vec{\beta}\cdot\vec{E}}{c}\vec{\beta}\right].
\end{equation}

As can be seen in Fig. \ref{fig:label:4a}, depending on the initial axion phases, the accumulated vertical EDM precession rates are different and random. However, this random axion phase issue can be resolved using two (or four) orthogonally set spin polarizations.  

An example of four orthogonally set polarizations are presented in Fig. \ref{fig:label:4c}. Fig. \ref{fig:label:4b} shows the effect of four orthogonally set spin polarizations on the $\omega_{EDM}$ for the case of $\varphi_{ax} =\pi/8$. From the measurement of individual spin polarization, one can calculate the total EDM precession rate using the relationship, $\omega_{EDM}=\sqrt{\omega_{EDM,S1}^{2}+\omega_{EDM,S2}^{2}}$, where $\omega_{EDM,S1}$ and $\omega_{EDM,S2}$ are two orthogonally set polarizations. The corresponding axion phase can be obtained by $\varphi_{ax}=arctan(\frac{\omega_{EDM,S2}}{\omega_{EDM,S1}})$.
As can be seen in the example shown in Fig. \ref{fig:label:4b}, the S1 and S3 states have large precession rates ($3.2\times10^{-7}$ rad/s) for the axion phase of $\varphi_{ax}=\pi/8$, and the precession directions are opposite to each other. On the other hand, note that the other two polarizations S2 and S4 have smaller $\omega_{EDM}$ than the two counter parts of the polarizations (S1, S3). In any case, the actual precession rate $\omega_{EDM}$ can be calculated using the formula shown above.

\begin{figure}
	\includegraphics[width=\linewidth]{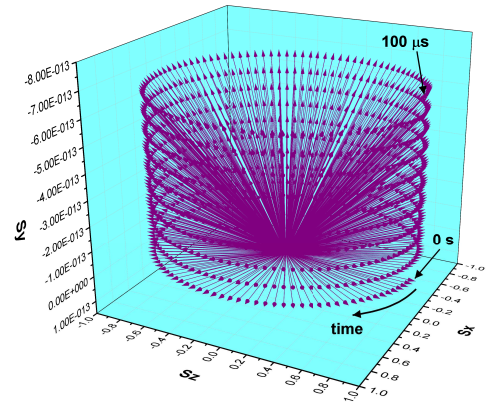}
	\caption{Deuteron spin precession in E/B combined ring.}
	\label{fig5}
\end{figure}

Fig. \ref{fig5} shows a simulation result for deuteron spin precession in a E/B combined ring with 100 kHz of g-2 frequency. The electric and magnetic fields used in the simulation were $7.05\times 10^{6}~\rm{V/m}$ and 0.38 T, respectively. The initial spin direction was set to the +z direction (0,0,1) and the total precession time shown in the figure is $100~\rm{\mu s}$. As can be seen, the vertical spin component ($S_y$) is accumulated while the horizontal spin precession takes place at the g-2 frequency. As mentioned before, the vertical precession rate depends on the initial axion phase.

The $\theta_{QCD}$ for a proton is reported to be three times larger than that of the deuteron \cite{deuteronEDMproposalToBNL,protonEDMproposalToBNL}. This means that the proton EDM described by Eq.  (\ref{eq1}) is three times larger than the deuteron case. However, in this sensitivity estimation, we used the same EDM $d$ for both deuteron and proton.

\section{Scanning method}

\begin{figure}
	\includegraphics[width=\linewidth]{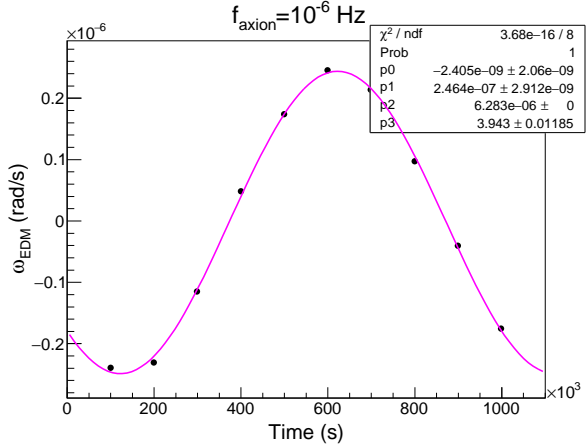}
	\caption{$\omega_{EDM}(t)$ measurement with frozen spin condition. The data is fit to the sine function for the sensitivity calculation.}
	\label{fig6}
\end{figure}

The sensitivities presented in Table \ref{table1} and \ref{table2} are based on the assumption that we know the axion mass (frequency) and perform the measurement at the same frequency for $8\times 10^7$ s. However, since we don't know the axion mass yet, all the possible frequency ranges have to be searched for. We used different scan methods for different frequency regions.  

For very low frequencies, $f_{ax}<100~\rm{\mu Hz}$, one can use the frozen spin method, which was used for the static EDM search, parasitically. With this method, one can take data repeatedly from the frozen spin condition for a limited storage time. The storage time has to be smaller than the axion coherence time and spin coherence time. For example, assuming a spin coherence time of $10^4$ s, one can use $10^4$ s as the maximum storage time, and with this storage time one can test up to $100~\rm{\mu Hz}$. 

Fig. \ref{fig6} shows an example of simulation results for an axion frequency of 1 $\mu$Hz. Each point corresponds to $\omega_{EDM}(t)$ whose data was taken for $10^4$ s. The data points are fit to the sine function and sensitivity is calculated from the fit results. The sensitivity calculated with this method was about $10^{-31} ~e\cdot$cm for frequencies $< 100~\mu$Hz. This method can be used for higher frequencies up to mHz if one uses a shorter measurement time ($<10^3$ s). No extra measurement is required for the axion search in this frequency region. One can do the static EDM experiment with frozen spin conditions and do the analysis for axion signal search afterwards. 

For high frequencies, for example $>$1 MHz, the resonance method can be used. Each run is done at a fixed frequency with the resonance conditions. In this case, the axion coherence time is used as the measurement time for each storage time. We used half of the axion width, $\Delta f_{ax}/2$, as the scan step and $\Delta f_{ax}=f_{ax}/Q_{ax}$. Assuming measurement time to be one axion coherent time for each frequency and $10^{11}$ particles per storage, $SCT=10^{4}$ s and $Q_{ax}=3\times 10^{6}$, the sensitivities were calculated to be about $10^{-26}-10^{-25}~e\cdot$cm. The total scan time can also be estimated using the measurement time for each frequency multiplied by the number of steps. For example, total scan steps for the frequency range of 1-100 MHz was $\sim 2.7\times 10^{7}$ and total scan time was $\sim 8\times 10^{7}$ s which corresponds to about 4 years of measurement time. For this estimation, we assumed the minimum measurement time to be a machine cycle time of 3 s.

\section{Axion gluon coupled EDM searches and the sensitivity of the experiments}

The current experimental limit comes from the ultra cold neutron trap method (nEDM) \cite{PhysRevX.7.041034, GREEN1998381,PhysRevLett.82.904,PhysRevLett.97.131801,BAKER2014184}. One can compare the sensitivities between the nEDM method and the storage ring EDM method. For example, the statistical error of the nEDM for one day's measurement is reported to be $6\times 10^{-25}~e\cdot$cm (see \cite{PhysRevLett.82.904} for details). This result is based on the experimental parameters such as the number of particles per storage (13,000 neutrons), free precession time of 130 s and electric field 450 kV/m. Compared with the proposed storage ring EDM method, the number of particles can reach up to $10^{11}$ per storage, the polarimeter efficiency can be about 2\% for the proton or deuteron case, the effective electric field can be over 1 GV/m, and the measurement time, which is limited by the spin coherence time, can be more than $10^{4}$ s. For example, for the $10^5$ Hz deuteron case with a measurement time of one day, the sensitivity is estimated to be $\sim 10^{-28}~e\cdot$cm. From this comparison, one can tell the storage ring method is more sensitive than the ultra-cold neutron trap method, by roughly more than 3 orders of magnitude. 

\begin{figure}
	\includegraphics[width=\linewidth]{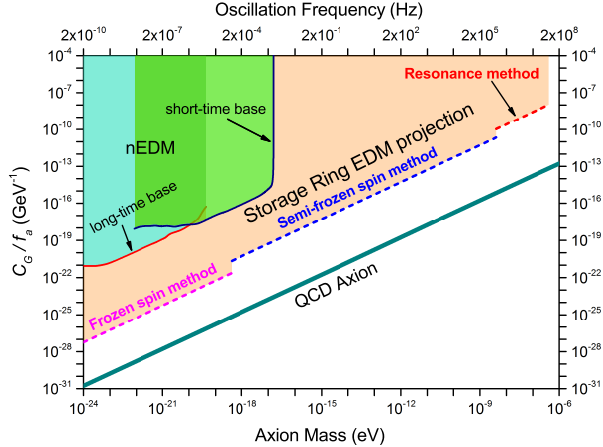}
	\caption{Experimental limits for the axion-gluon coupled oscillating EDM measurement. The nEDM results are included for comparison. See reference  \cite{PhysRevX.7.041034} for detailed nEDM results. The limits for three different frequency regions are indicated with different colors of broken lines. Average values are used for each frequency region. See the column for $SCT=10^{4}$ s and $Q_{ax1}=3\times 10^{6}$ in Table \ref{table1} and \ref{table2} for the relevant numbers. If $10^{10}$ is used as the axion quality factor, the $C_G/f_a$ can be improved by 1 or 2 orders of magnitude in the high frequency region. Estimation of the limit plot for storage ring EDM method is made by assuming that the axion mass (frequency) is known and the measurement is performed for $8 \times 10^7$ s at that frequency. However, for very low frequency ranges ($ f_{ax}<$ 1 mHz), the axions can be searched for with high sensitivity using the frozen spin method without knowing the axion frequency. See the text for details.}
	\label{fig7}
\end{figure}

Fig. \ref{fig7} shows the experimental limits for the axion-gluon coupled oscillating EDM measurements. For comparison, we include the nEDM results, which were taken from Fig. 4 in reference \cite{PhysRevX.7.041034}. Based on the calculation results shown in Tables \ref{table1} and \ref{table2}, the projected limit of the storage ring EDM method is drawn by the red dotted line for the resonance method (1 MHz $< f_{ax} <$ 100 MHz), the blue dotted line for the semi-frozen spin method (100 $\mu$Hz$< f_{a} <$1 MHz) and the magenta dotted line for the frozen spin method ($f_{ax}<100~\mu$Hz). For an axion frequency below $100~\mu \rm{Hz}$, one can take data from the frozen spin condition which is used for the static EDM experiment. By combining many separate runs of data, as the ultra-cold neutron EDM experiment did (periodogram analysis) \cite{PhysRevX.7.041034}, searching for the very low frequency region is possible. Even lower frequency searches are possible with data collected for longer times. The sensitivities used for the projected limit plot were averages for the $SCT=10^{4}$ s and $Q_{ax1}=3\times 10^{6}$ in Table \ref{table1}, and Table \ref{table2} for the 100 MHz proton case.

The CASPEr experiment proposes cosmic axion searches using the nuclear magnetic resonance (NMR) method \cite{PhysRevX.4.021030,Garcon:2017ixh}. They utilized the resonance of nuclei spin precession with the oscillating axion field. With the NMR method, one can search the high frequency region when a strong magnetic field is used. 

The axion field is proportional to the square root of the local axion dark matter density. For this calculation, we used $\rho_{DM}^{local}\approx 0.3~\rm{GeV/cm^{3}}$ as the local cold dark matter density. 
Furthermore, it can take advantage of the recently proposed local dark matter enhancement factors from the focusing effects due to planetary motion \cite{Zioutas:2017klh,BERTOLUCCI201713}.
 
\section{Summary and conclusion}
 As a candidate for dark matter, the axion has been the target of extensive searches  using microwave cavities and other methods. The fact that the axion-gluon coupling can produce an oscillating EDM in nucleons led to the novel idea of measuring the oscillating EDM in hadronic particles like the proton and deuteron. We propose using the storage ring technique to measure the axion induced oscillating EDM at the resonance conditions between the axion frequency and $g-2$ spin precession frequency. 
 
 In this study, we calculated the electric field and magnetic field that are required for the resonance conditions. With the experimental conditions, we estimated the achievable sensitivities, and the result shows the experiment is more sensitive than the planned static EDM measurement ($10^{-29} ~e\cdot$cm) by at least one order of magnitude, $\leq10^{-30}~e\cdot$cm. This sensitivity is achieved if we assume that we know the axion frequency and spend all the experimental time at one frequency value. At very low frequencies, $f_{ax}<1$ mHz, one can search for the axion with the sensitivities of $10^{-31}-10^{-32}~e\cdot$cm using the frozen spin method, without knowing the axion frequencies.

A wide range of frequencies ($10^{-9}$ Hz $-$ 100 MHz) of axion dark matter can be searched by using both deuterons and protons in the same storage ring. Even though the proposed method does not reach the estimated sensitivity needed to reach the theoretical models of axion dark matter induced oscillating EDM, it promises to be one of the most sensitive ways to look for axions over a wide frequency range.  

\section*{Acknowledgement}
This work was supported by IBS-R017-D1-2018-a00 of the Republic of Korea. This idea was developed when one of the authors (YkS) was invited to give a talk at Stanford University in 2013, by Peter Graham, Surjeet Rajendran, and Savas Dimopoulos.  They suggested to him that the oscillating theta idea might be applicable to the storage ring EDM method and we thank them for it.



%% References

%\section*{References}
\bibliographystyle{unsrt}
\bibliography{reference}






% \end{thebibliography}

\end{document}